\documentclass[aps,prb,twocolumn,showpacs,amssymb]{revtex4}

\usepackage[dvips]{graphicx}
\usepackage{dcolumn}
\usepackage{bm}
\def\be{\begin{equation}}
\def\en{\end{equation}}

\def\p{\partial} 
\newcommand{\av}[1]{\langle{#1}\rangle}
\def\gs{\gtrsim}

\newcommand{\bi}[1]{\mbox{\boldmath$#1$}}
\def\p{\partial}
\def\bea{\begin{eqnarray}}
\def\ena{\end{eqnarray}}

\renewcommand{\theequation}{\arabic{section}.\arabic{equation}}

\def\hbk{{\hat{\bi k}}}

\begin{document}


\title{Phase transition in  compressible Ising systems  
at fixed volume }


\author{Akira Onuki and Akihiko Minami}


\affiliation{Department of Physics, Kyoto University, Kyoto 606-8502,
Japan}


\date{\today}

\begin{abstract}
Using a Ginzburg-Landau model, we study the phase transition behavior 
of compressible Ising systems 
at constant volume by varying  the temperature $T$ and 
the applied magnetic field $h$. 
We show that two phases can coexist macroscopically 
in equilibrium within a closed region in the $T$-$h$ plane. 
It occurence is favored  near  tricriticality.  
We find a  field-induced critical point, where 
the correlation length  diverges,  
the difference of the coexisting two phases 
 and the surface tension vanish, but the isothermal magnetic 
susceptibility does not diverge  in the mean field theory. 
We also investigate phase ordering numerically. 
\end{abstract}


\maketitle


\section{Introduction}

 Solids are 
under the influence of elastic constraints 
and their phase transitions are often decisively 
influenced by couplings of the order parameter and 
the elastic field \cite{Onukibook}. Such 
 elastic effects strongly depend on the nature of the 
coupling  and their understanding 
is crucial  in technology.  In the present work, we will 
focus on  the phase transition behavior of 
compressible ferromagnets 
or antiferromagnets, which has long been studied  
theoretically in the physics community  
\cite{Larkin,Baker,Sak,Imry,Wegner,Barber,Lubensky,Halperin,La1,La3,La4}. 
 In real materials, the short-range 
 spin interactions depend  on the distances  among 
the spins, so the spin fluctuations are coupled 
to the elastic dilation strain. 
In the literature on this problem, the main issue has been  
the effect of  the elastic coupling 
on the critical behavior of the spin system.  
A remarkable but subtle result of  
 the renormalization group calculations 
\cite{Lubensky,Halperin}  is   that 
the cubic elastic anisotropy becomes 
increasingly important  on 
approaching the critical point 
(which is determined in the absence of the anisotropy). 
This renormalization effect should  trigger   
a first order phase transition 
sufficiently close to the critical point. 
Simulations have been 
prerformed on  compressible  Ising systems  
 and a number of 
numerical results  still remain not well understood 
 \cite{La1,La3,La4}. 
These theories and simulations show that 
 the  phase  transition   depends  
on whether the pressure or the volume is fixed.

In this paper, we will present a mean field theory 
of compressible Ising systems at constant volume 
using  a Ginzburg-Landau free energy. 
Our main objectives are 
to  demonstrate the presence of unique  
two phase coexistence  near the tricritical point 
and to examine phase ordering after 
changing the temperature.  
Though our  theory is 
a rough approximation, it will  provide 
 overall phase behavior 
 for general values of the parameters.

The organization of this paper is as follows. 
In  Sec. II, we will present 
a model,  in which the order parameter and the elastic field are  
coupled, and eliminate the elastic degrees of freedom 
assuming the mechanical equilibrium condition. 
In Sec. III, we will examine the phase behavior 
in the plane of the temperature $T$ and the ordering field $h$. 
Detailed calculations will also be given on  
the susceptibility, the correlation length,   
and the surface tension.  
The presence of a unique field-induced critical point 
will also be reported. In Sec. IV, we will numerically 
integrate  the  time-dependent Ginzburg-Landau equation 
in two dimensions (2D). 
In the appendix, we will derive 
 the free energy at  constant pressure (or applied 
stress), where  two-phase coexistence can be realized  only 
on lines in the $T$-$h$ plane.

\section{Theoretical background}
\subsection{Ginzburg-Landau free energy}
\setcounter{equation}{0}

We assume that   a single-component   order parameter  
$\psi$ is coupled to  the elastic displacement $\bi u$. 
We set up the Ginzburg-Landau free energy functional 
$F=F\{\psi,{\bi u}\}$ in the form \cite{comment},  
\be 
F= \int d{\bi r} \bigg [f_0 +
\frac{C}{2} |\nabla\psi|^2  + \alpha \psi^2e_1+  
f_{el}\bigg ], 
\en  
where the space integral is within the system with volume $V$. 
The first part $f_0=f_0(\psi)$ depends on $\psi$    as  
\be 
{f_0}= 
\frac{ \tau}{2} \psi^2 +
\frac{{\bar u}}{4} \psi^4+
\frac{v}{6} \psi^6 -h\psi. 
\en 
The coefficient  $ \tau$ depends on the temperature $T$ as 
\be 
{ \tau}=A_0 (T-T_0), 
\en 
where  $A_0$ is a positive constant and $T_0$ is 
the  critical temperature in the absence 
of the elastic coupling. The other coefficients are treated 
to be independent of $T$.  We fix the other field variables 
such as the hydrostatic pressure. 
The coefficients  $v$ and $C$ are  positive, 
while $\bar u$ can be  either positive or negative. The 
 $h$ represents a   magnetic or electric 
field  conjugate to $\psi$.    For  antiferromagnetic 
materials, no uniform field conjugate to the 
antiferromagnetic order can be realized, so $h=0$. 
We may assume $h \ge 0$ without loss of generality. 
If $h=0$,  $F$ is invariant with respect to 
$\psi\rightarrow -\psi$. 
The $\alpha$ represents  the strength of the 
coupling between $\psi^2$ and 
 the dilation strain,     
\be 
e_1 = \nabla\cdot{\bi u}.   
\en 
This coupling arises when the 
interaction among the  fluctuations of $\psi$ 
 depends  on   the local 
lattice expansion or contraction.

In cubic crystals, 
 the   elastic energy density is of  the form,    
\be 
f_{\rm el}= \frac{C_{11}}{2}  \sum_{i}\epsilon_{ii}^2+
 \sum_{i\neq j} \bigg [ \frac{C_{12}}{2}
\epsilon_{ii}\epsilon_{jj}  
+C_{44} \epsilon_{ij}^2 \bigg ],
\en 
where  $C_{11}$, $C_{12}$, and $C_{44}$ are the usual elastic 
moduli assumed to be constant, and  
$\epsilon_{ij}= (\nabla_i u_j + \nabla_j u_i)/2$ 
is the symmetrized strain tensor. 
The dependence of the elastic moduli on 
$\psi^2$ can be important at low temperatures, however. 
 Hereafter $\nabla_i= \p/\p x_i$.  
The  elastic stress tensor $\sigma_{ij}$ is  expressed as 
\bea
\sigma_{ii}&=&(C_{11}-C_{12})\epsilon_{ii} 
+ C_{12}e_1 + \alpha \psi^2, \nonumber \\
\sigma_{ij}&=&2C_{44} \epsilon_{ij} \quad\qquad\quad (i \neq j).
\label{eq:10.2.2}
\ena   
Nonvanishing $\psi^2$ gives rise to 
a  change  in the  diagonal stress components. 
We then obtain 
$
\sum_j\nabla_j\sigma_{ij}= - {\delta F}/{\delta u_i}$,  
where $\psi$ is fixed in the functional 
derivative of $F$ with respect to $u_i$.  
Note that a constant hydrostatic pressure $p_0$ 
can be present in the reference state, where 
the total stress  tensor  is   $p_0\delta_{ij}-  
\sigma_{ij}$.

\subsection{Elimination of elastic field at fixed volume}

The elastic field $\bi u$ is determined by $\psi$ 
under  the mechanical equilibrium condition, 
\be 
\sum_j\nabla_j\sigma_{ij}=0.
\en 
Furthermore,  in this paper, 
we  impose the periodic boundary condition 
on  $\delta {u}$ in 
 the region $0<x, y, z<V^{1/d}$. 
This can be justified when the solid boundary is  mechanically  
clamped. See Appendix A for the  case of fixed applied pressure. 
The space averages of the strains then vanish; 
for example, $\av{e_1}=0$. Hereafter $\av{\cdots}= \int d{\bi r}
(\cdots)/V$. The following procedure of eliminating the 
elastic field has been derived by many authors in the literature in 
physics and engineering \cite{Onukibook,Larkin,Sak,Imry,La3,Kha}.

It is  convenient to   use  the Fourier 
transformation,  
$
 u_j({\bi r})= 
\sum_{\bi k} u_{j{\bi k}} \exp({i{\bi k}\cdot{\bi r}}),   
$  
where $\bi k$ is the wave vector. 
Then the  Fourier component of $e_1$ 
is expressed as  
\be 
{e_{1{\bi k}}}
= -   {\alpha}\varphi_{\bi k}/{[C_{12}+ C_{44}+ C_{44}\zeta(\hbk)] }, 
\en 
where $\varphi_{\bi k}$ is the Fourier 
component of the  variable, 
\be 
\varphi({\bi r})= \psi^2 -\av{\psi^2}.
\en 
The space average of  $\varphi$ is made to vanish. 
The  $\zeta(\hbk)$ is a function of 
the direction of the wave vector $\hbk=k^{-1}{\bi k}$ 
and is defined by  
\be
\zeta(\hbk)^{-1} =  
\sum_j {\hat k_j^2}/({{1+ \xi_{\rm a}\hat k_j^2}}),
\en 
where $\xi_{\rm a} $ is the degree of  cubic 
anisotropy, 
\be
\xi_{\rm a} =( C_{11}-C_{12})/C_{44}-2.
\en 
We  have  $\zeta(\hbk)=1$  in the isotropic elasticity   
$\xi_{\rm a}=0$. 
After some calculations, 
we may eliminate $\bi u$ in $F$ 
to obtain  the free energy $F=F\{\psi\}$ 
of  $\psi$ only in   the form \cite{Onukibook}, 
\be
{F} =  \int d{\bi r}\bigg [ f_0 
+\frac{C}{2} |\nabla\psi|^2 \bigg]- 
 \frac{1}{2V}\sum_{\bi k} w({\hat{\bi k}})  |\varphi_{\bi k}|^2   .
\en  
The second term on the right hand side arises 
from the elastic coupling and is negative, where    
\be 
w(\hbk)=  \alpha^2 /{[C_{12}+ C_{44}+ C_{44}\zeta(\hbk)] }.  
\en 
The functional derivative of $F$ is performed to give 
\be 
\frac{\delta F}{\delta \psi}= 
f_0' -C\nabla^2\psi + 2\alpha e_1\psi,
\en 
where $f_0'= \p f_0/\p \psi$ and 
the Fourier transformation of $e_1$ is  in Eq.(2.8). 
In equilibrium we require 
$\delta F/\delta\psi=0$.

We further simplify our free energy. 
In the isotropic elasticity,  
$w(\hbk)$ is a constant independent of $\hbk$ and 
$
e_1= -\alpha\varphi /C_{11} .
$ 
Then $F$ is rewritten as     
\be
{F} =  \int d{\bi r}\bigg [ f_0 
+\frac{C}{2} |\nabla\psi|^2 - 
\frac{\beta}{4} (\psi^2 -\av{\psi^2})^2  \bigg] ,
\en  
where $\varphi$ is explicitly 
written in terms of $\psi$ and 
 $\beta$ is a positive constant defined  by 
\be 
\beta =2\alpha^2 /C_{11}.
\en  
The presence of the space average 
$\av{\psi^2}$ is a unique aspect arising from elasticity.

In cubic solids with $\xi_{\rm a}<0$, 
$w(\hbk)$ is maximized  
along one of the principal 
crystal axes (say, along the $[100]$ direction in 3D) \cite{Onukibook}. 
If $\xi_a>0$,  it is maximized 
for $\hat{k}_j^2=1/d$ 
for all $j$  (say, along $
 [111]$ in 3D).  Let $w_M$ be the maximum  of $w(\hbk)$ 
 attained along these soft directions; then, 
\bea 
w_M &=& \alpha^2/C_{11} \qquad\quad  (\xi_a<0),\nonumber\\ 
&=& \alpha^2/[K + (2-2/d)C_{44}]\quad (\xi_a>0).
\ena  
where $K= C_{11}/d + C_{12}(1-1/d)$ 
is the bulk modulus.   In 2D,  
$w(\theta)=w(\hat{\bi k})$ is a periodic 
function of the angle $\theta$ 
defined by $k_x/k=\cos\theta$ and  
$k_y/k=\sin\theta$ with period $\pi/2$, 
as displayed in Fig. 1. 
In phase ordering processes,  
the interface normals tend to be parallel 
to these soft directions, resulting in cuboidal domains 
\cite{Onukibook,comment,Onuki_cubic,Nie}. 
If the spatial inhomogeneity is 
mostly along these soft directions except for the edge 
regions of the domains, the free energy is approximately 
 given by Eq.(2.15) with 
\be 
\beta = 2w_M. 
\en

\begin{figure}[htbp]
 \includegraphics[scale=0.4]{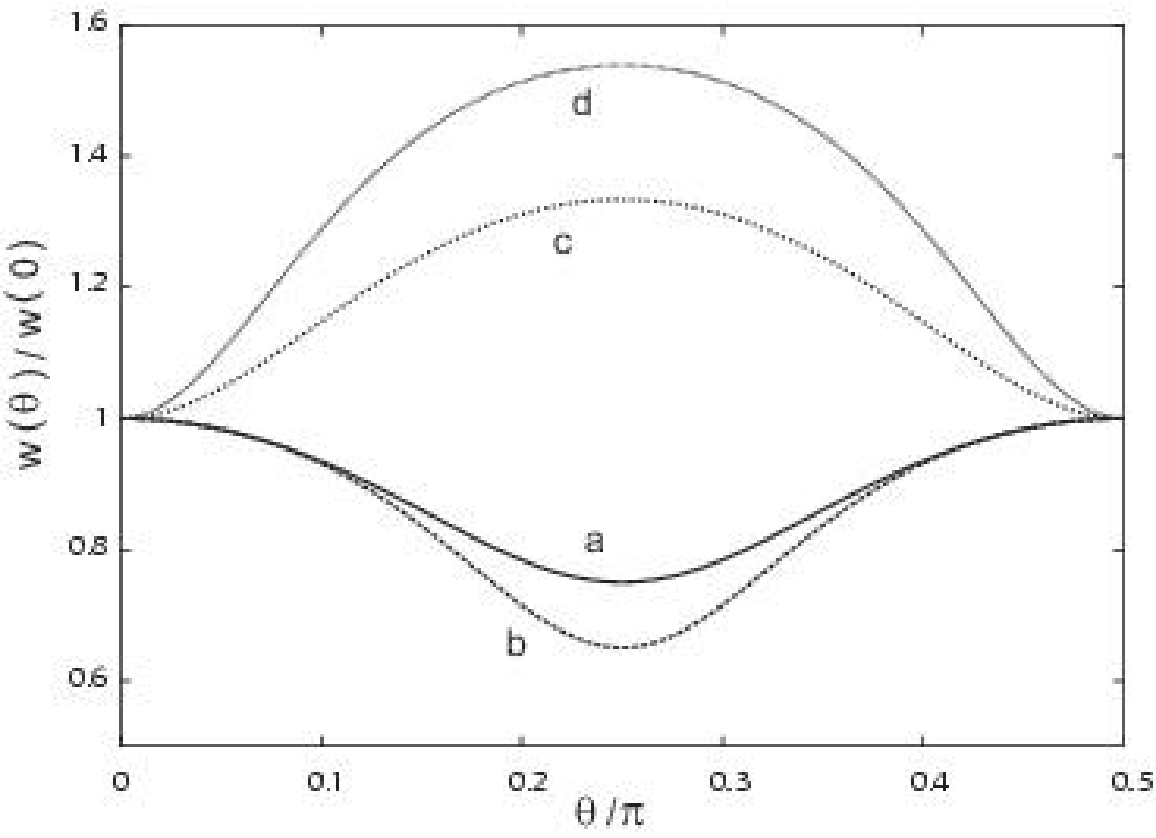}
 \caption{$w(\theta)/w(0)$ in 2D 
as a function of $\theta/\pi$ for $((C_{11}-C_{12})/2K, 
C_{44}/K)= (0.5,1)$ (a), $(0.3,1)$ (b), 
$(1,0.5)$ (c),  and  $(1,0.3)$ (d).  
The maximum of $w(\theta)$ 
is $w(0)$ for $\xi_a<0$ and $w(\pi/2)$ for $\xi_a>0.$
}
\end{figure}

\subsection{One phase states}

We  start with the  free energy Eq.(2.15). 
If the system consists of a single ordered phase 
in equilibrium, we have $\psi^2= \av{\psi^2}$ 
and the homogeneous $\psi$ is determined by 
\be 
f_0'= (\tau +\bar{u}\psi^2+v\psi^4)\psi-h=0,  
\en 
where the elastic coupling disappears.  The inverse 
susceptibility $\chi^{-1}=(\p h/\p \psi)_\tau$ 
is given by 
\be 
\chi^{-1}=\p^2 f_0/\p\psi^2= 
\tau +3\bar{u}\psi^2+5v\psi^4.
\en

We may consider the structure factor 
$S_k$ of the thermal fluctuations of 
the Fourier component $\psi_{\bi k}$ in the bulk region.  
To calculate it,  we superimpose 
plane wave fluctuations of $\psi$ on the homogeneous 
average. The increase of 
the free energy in the second order 
yields $S_k$  in the Ornstein-Zernike form 
\be 
S_k= 1/C(k^2+ \kappa^2), 
\en 
where  $\kappa$ is the inverse correlation determined by 
\bea 
C\kappa^2&= &\p^2 f_0/\p\psi^2-3\beta\psi^2 +\beta\av{\psi^2}
\nonumber\\
&=& \tau + (3\bar{u}-2\beta)\psi^2+5v\psi^4. 
\ena  
In the second line, 
  we have set  $ \av{\psi^2}=\psi^2$ 
because of  the existence of a single  phase only. 
Note that $C\kappa^2$ in the second line of Eq.(2.22) is smaller than 
$\chi^{-1}$ in Eq.(2.20)  by $2\beta\psi^2$. 
In cubic solids,   $\kappa$ represents the inverse correlation length 
for the fluctuations varying 
in the softest directions. 
Let $\tau$ take  a small negative value at $h=0$ 
in the case $\bar{u}>0$; then, 
$\psi^2\cong |\tau|/{\bar u}$ from Eq.(2.19), 
leading to  
$ 
C\kappa^2\cong 2(1-\beta/{\bar u})|\tau|
$
from Eq.(2.22). 
The positivity of $\kappa^2$ 
is  attained  only for $\beta<{\bar u}$. 
Obviously, the disordered phase with $\psi=0$ 
is unstable for $\tau<0$. 
The ordered phase with 
$\psi^2= -{\bar u}/2v+ \sqrt{{\bar u}^2/4v^2- \tau}$ 
(which is the solution of Eq.(2.19) at $h=0$) 
becomes unstable for $\tau>\tau_{\rm in }$. 
In particular,  as $h\rightarrow 0$, we find  
\be 
\lim_{h\rightarrow 0} \tau_{\rm in}= -(\beta^2-{\bar u}^2)/4v.  
\en 


\section{Two phase coexistence}
\setcounter{equation}{0}
\subsection{Two  phase states }

\begin{figure}[htbp]
 \includegraphics[scale=0.4]{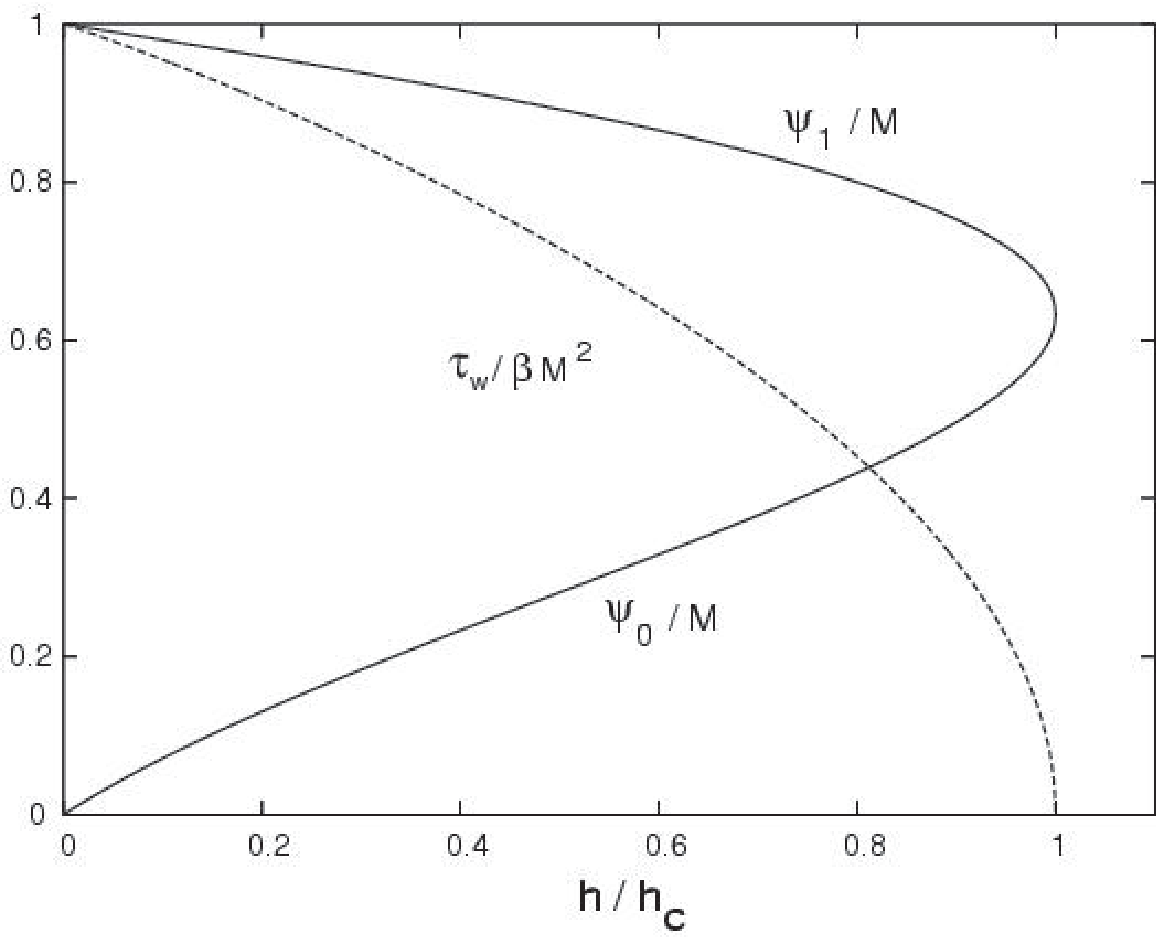}
 \caption{Normalized order parameters 
 $\psi_1/M$ and $\psi_0/M$ and 
normalized width of the temperature window 
$\tau_w/\beta M^2$ versus normalized field $h/h_c$ 
in  two phase coexistence, 
where $M$ and $h_c$ are  defined by (3.9) 
and (3.12), respectively.  }
\end{figure}

\begin{figure}[htbp]
 \includegraphics[scale=0.4]{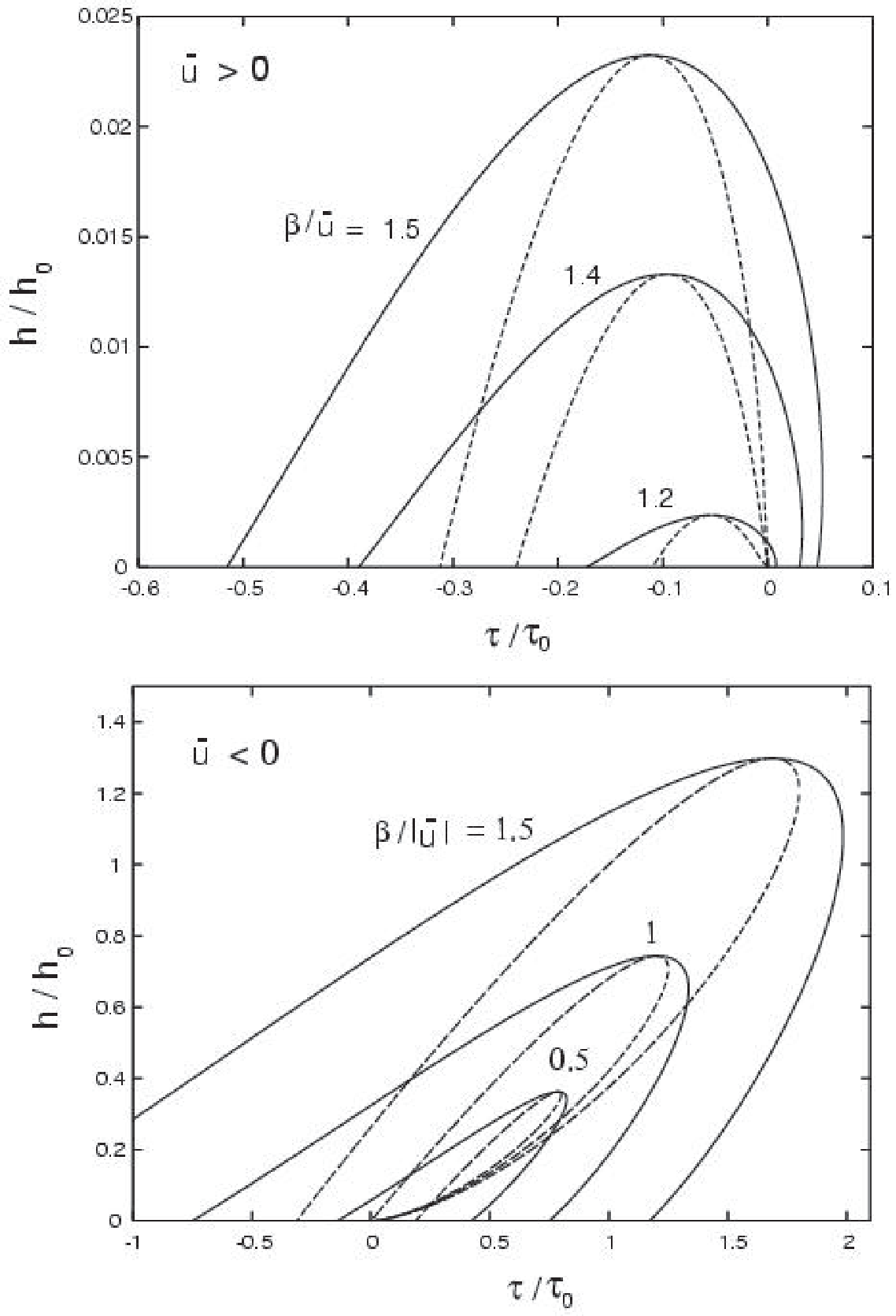}
 \caption{Phase diagrams  in the $\tau$-$h$ plane 
with ${\bar u}>0$ for $\beta/{\bar u}=1.5, 1.4$, and 1.2 
(upper plate) and  with 
${\bar u}<0$ for $\beta/|{\bar u}|=1.5, 1$, and 0.5  
(lower plate).  
The $\tau$ and $h$ are scaled by 
 $\tau_0= {\bar u}^2/v$ and 
 $h_0= v(|{\bar u}|/v)^{5/2}
= 
|6(\beta/{\bar u}-1)/5|^{5/2}h_c/12$, so 
$h/h_0$ is large around $h\sim h_c$ 
for negative $\bar u$.  
The system is in two phase coexistence 
inside   each coexistence curve (solid line), while it 
is in a one phase state outside it. 
Instability curve (dotted line)
 merges each coexistence curve at the critical point, 
inside  which one phase states 
are linearly unstable. @
}
\end{figure}

We show that   
two phases can coexist in a temperature 
window $\tau_c- \tau_w<\tau<\tau_c$ 
if the  parameter,   
\be 
u= {\bar u}-\beta, 
\en 
is negative \cite{Sak,Imry} and $h$ is smaller than a critical 
field $h_c$, where $\tau_c$, $\tau_w$, and 
$h_c$ will be determined below.    We of course  
have $u<0$ if  $ {\bar u}<0$  or if the system 
undergoes a first order phase transition even 
without the elastic coupling.  For $0\le h\le h_c$ 
the two phases are characterized 
by $\psi=\psi_0$ and $\psi_1$ 
with $\psi_1\ge \psi_0\ge 0$. 
As $h\rightarrow 0$ we have $\psi_0\rightarrow 0$, 
while as $h\rightarrow h_c$ 
we have  $\psi_1-\psi_0\rightarrow 0$. 
We will show that the space average $\av{\psi^2}$ 
in the free energy (2.15) gives rise to the two phase coexistence. 
If it were neglected, we would have the usual tricritial 
point at $\tau=u=0$ (see the last paragraph of this subsection) 
\cite{Onukibook, Griffiths_tri}.

If the volume fraction of the phase with 
$\psi=\psi_1$ is written as $\phi$, 
we have 
\be 
\av{\psi^2}= \phi \psi_1^2+ (1-\phi)\psi_0^2.
\en  
The average free energy density 
$\av{f}=F/V$ is given by 
\bea 
\av{f} &=& \phi f_0(\psi_1)+ (1-\phi)f_0(\psi_0) \nonumber\\
&& 
+\frac{1}{4}\beta(\psi_1^2-\psi_0^2)^2(\phi^2-\phi) .
\ena 
Here the interface free energy is neglected. 
The  minimization conditions 
of  $\av{f}$ with respect to 
$\psi_1$ and  $\psi_0$  are given by 
\bea 
f_0'(\psi_1) &-&\beta (1-\phi)(\psi_1^2-\psi_0^2)\psi_1=0,\\
 f_0'(\psi_0) &+&\beta\phi(\psi_1^2-\psi_0^2)\psi_0=0,
\ena 
which  are equivalent to $\delta F/\delta\psi=0$ 
at $\psi=\psi_1$ and $\psi_0$. 
We also minimize $\av{f}$ with respect to  $\phi$ to obtain 
\be 
f_0(\psi_1) -f_0(\psi_0) +\frac{\beta}{4}
(\psi_1^2-\psi_0^2)^2(2\phi-1)=0,
\en 
which means that the two phases 
have the same free energy density. 
Note that the quadratic term $(\propto \phi^2)$ 
in $\av{f}$ in Eq.(3.3) is positive for 
$\psi_1-\psi_0>0$.  Thus, 
 for small $f_0(\psi_1) -f_0(\psi_0) $, 
 a minimum of $\av{f}$ can be attained 
as a function of $\phi$ in the range $[0.1]$.  
These equations may be solved 
for the simple free energy density (2.2). 
By eliminating $\phi$ we   derive 
the equations for 
$\psi_1$ and $\psi_0$ as 
\bea 
h/v &=&  \psi_1\psi_0 (\psi_0+\psi_1)^3/3, \\
-u/v&=&  \psi_1^2+\psi_0^2+ \frac{1}{3} (\psi_0+\psi_1)^2
\ena 
where $u$ is defined by Eq.(3.1).
The  negativity of $u$ is required by  Eq.(3.8). 
Thus  $\psi_1$ and $\psi_0$ are independent of $\tau$. 
 As $h\rightarrow 0$, we have $\psi_0=0$ and 
$\psi_1=M$, where 
\be 
M= (3|u|/4v)^{1/2}.
\en  
It is convenient to  express  $\psi_1$ and $\psi_0$ as 
\be 
\psi_1= \frac{q}{2}+ \sqrt{\frac{q^2}{4}- \frac{3h}{vq^3}}, \quad 
\psi_0= \frac{q}{2}- \sqrt{\frac{q^2}{4}- \frac{3h}{vq^3}},
\en 
where $q$ satisfies 
\be 
h= \frac{2v}{9}q^3(q^2-M^2) .
\en 
Then $q/M$  is a dimensionless function of $h/vM^5$, 
tending to unity as $h\rightarrow 0$. 
 The difference 
$\psi_1-\psi_0= (q^2- 12h/vq^3)^{1/2}$ decreases  with increasing $h$.  
A  field-induced criticality is attained for 
$h=h_c$  and  $\tau=\tau_c$, where  
\bea 
h_c &=& (8/5)^{5/2}vM^5/12,\\
\tau_c &=& 4vM^4/5 - 2\beta M^2/5.
\ena  
The critical value of the order parameter is 
\be 
\psi_c = (2/5)^{1/2}M= (3|u|/10v)^{1/2}.
\en 
For small positive $h_c-h$ we obtain 
\be 
\psi_1-\psi_0 \cong \frac{2}{5} M(1-h/h_c)^{1/2}. 
\en 
For $h>h_c$ we have a unique one phase state 
where $\psi$ is determined 
by Eq.(2.24).  In Fig. 2,  we show $\psi_1/M$ and 
$\psi_0/M$ versus $h/h_c$.

Next the  volume fraction  of the more 
ordered phase $\phi$  is calculated.  
From  Eq.(3.5) it 
depends on $\tau$ as 
\be 
\phi= (\tau_{cx}-\tau)/ \tau_w.
\en 
This relation holds   for  $\beta>{\bar u}$  and  $h <h_c$ with  
\bea 
\tau_{cx}&=& -{\bar u}\psi_0^2-v\psi_0^4 + 
\frac{v}{3}\psi_1(\psi_0+\psi_1)^3,\\
\tau_w&=&\beta (\psi_1^2-\psi_0^2). 
\ena 
In Fig. 2, the normalized window 
width $\tau_w/\beta M^2$ is also displayed  as a 
function of $h/h_c$.    
Since $\phi$ is in the range $0< \phi< 1$, 
the two-phase coexistence is realized in the window region, 
\be 
\tau_{cx}-\tau_w <\tau<\tau_{cx}.
\en 
For  $\tau$ below $\tau_{cx}$ 
the more ordered phase starts to appear,  and  $\tau_w$ 
is the width of the temperature window. 
As $h\rightarrow 0$,  $\tau_{cx}$ and $\tau_w$ 
tend to the following values, 
\bea 
\lim_{h\rightarrow 0}\tau_{cx}&=&{v} M^4/3= {3u^2}/{16v},\\
\lim_{h\rightarrow 0}\tau_{w}&=&{\beta} M^2=
 3\beta(\beta-{\bar u})/{4v}. 
\ena 
 On the other hand, 
as $h\rightarrow h_c$, 
the upper and lower bounds in Eq.(3.19) 
meet at $\tau=\tau_c$ and 
behave as 
 $\tau_{cx}  \cong 
\tau_c+\beta\psi_c (\psi_1-\psi_0)$ 
and $\tau_{cx} -\tau_w   \cong 
\tau_c-\beta\psi_c (\psi_1-\psi_0)$, where 
$\psi_1-\psi_0$  depends on $h_c-h$ as in  Eq. (3.15). 
In Fig. 3, we show  the phase diagrams  in 
the $\tau$-$h$ plane for $\bar{u}>0$ 
and  for $\bar{u}<0$,  separately,  where  
the coexisting curves,  $\tau=\tau_{cx}$ and 
$\tau=\tau_{cx}-\tau_w$,  and the instability curves 
are displayed.  The latter are  obtained 
by setting  $C\kappa^2=0$ in Eq.(2.22) 
using $\psi$ determined by  Eq.(2.19) (see the discussions 
above Eq.(2.23)). 
These   curves meet at the corresponding critical 
point $h=h_c$ and $\tau=\tau_c$ 
given by Eqs.(3.12) and (3.13).

The usual theory of tricriticality  
\cite{Onukibook,Griffiths_tri,Gammon} 
starts with the free energy density, 
\be  
f= \frac{\tau}{2}  \psi^2+ \frac{u}{4} 
\psi^4+ \frac{v}{6} \psi^6-h\psi, 
\en  
for systems with short-range interactions. 
For this  model  a first order phase transition 
line \cite{comment2}  appears in the $\tau$-$h$ plane for $u<0$. 
(i) The  line starts from the $\tau$ axis 
($h=0$) at the transition point  given by    $\tau={3u^2}/{16v}$ 
where  $\psi^2= 3|u|/4v$  
in the emerging ordered phase. 
These values coincide with 
those in  Eqs.(3.20) and  (3.9) in our elastic model.
(ii) The line 
ends at a  field-induced critical point, where  
$\psi^2= 3|u|/10v$,  $h=8v(3|u|/10v)^{5/2}/3$, and $\tau= 9u^2/20v$. 
The   critical values of $\psi$ and $h$ 
 coincide  with those in Eqs.(3.14) and (3.12). 
However,  the  critical 
value of $\tau$ 
is higher   than that  in Eq.(3.13) by $2\beta M^2/5$.

\subsection{Magnetization,  susceptibility, and specific heat}

\begin{figure}[htbp]
 \includegraphics[scale=0.4]{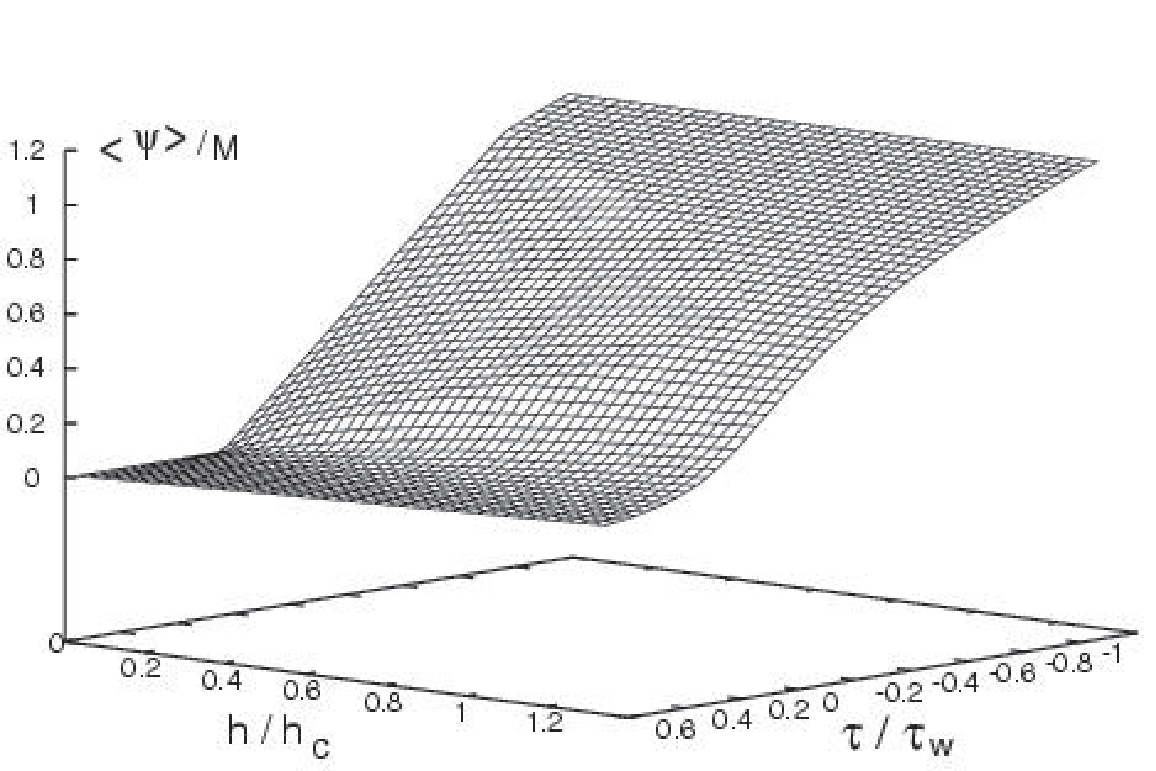}
 \caption{Normalized average magnetization 
$\av{\psi}/M$ as a function of $h/h_c$ and $\tau/\tau_w$ 
for $\beta/\bar{u}=1.2$  calculated from Eqs.(2.19) and (3.23), 
where $M$, $h_c$, and $\tau_w$ are defined by Eqs.(3.9), 
(3.12), and  (3.18), respectively. }
\end{figure}

\begin{figure}[htbp]
 \includegraphics[scale=0.4]{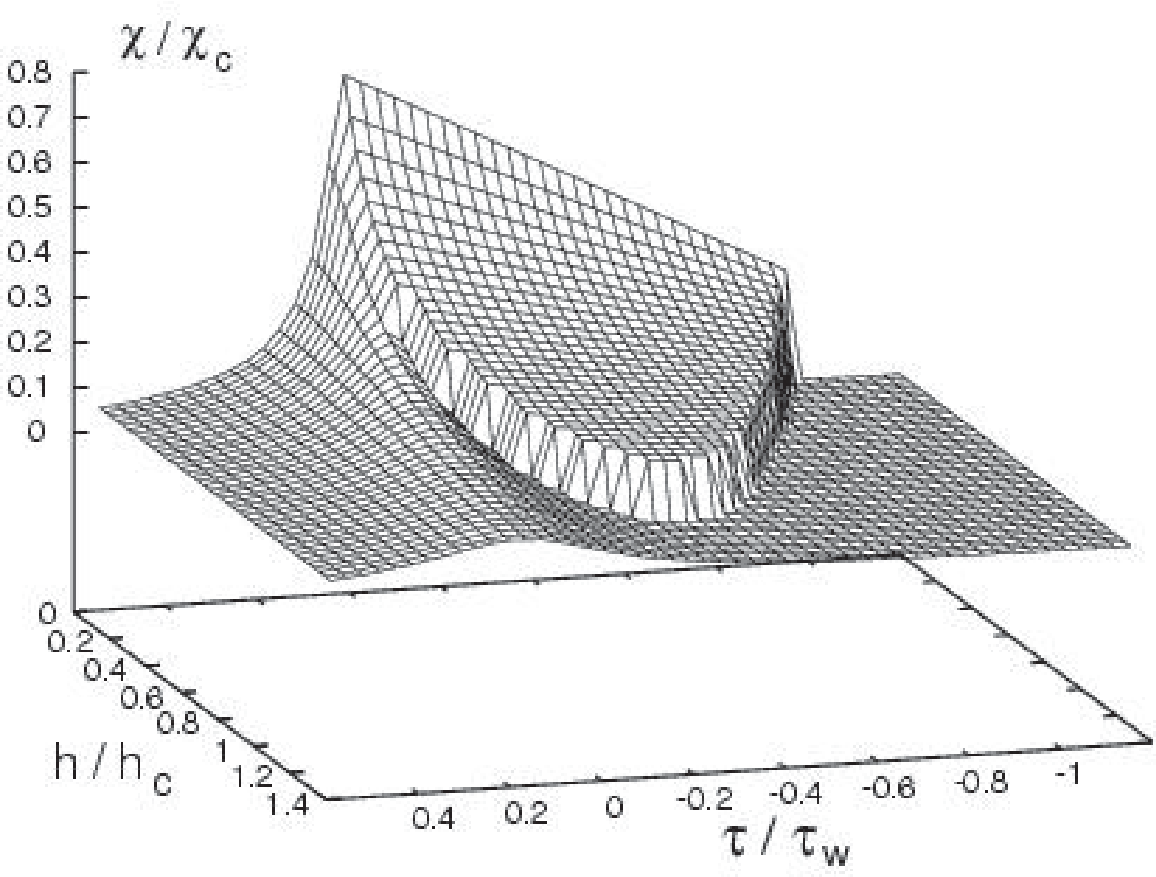}
 \caption{ Normalized susceptibility 
$\chi/\chi_c$ as a function of $h/h_c$ and $\tau/\tau_w$ 
for $\beta/\bar{u}=1.2$, where $\chi_c= M/h_c$.  It is 
calculated from Eqs.(2.20) and (3.24).  It 
increases  discontinuously at the phase boundary 
from the one phase 
region to the two phase region.}
\end{figure}

\begin{figure}[htbp]
 \includegraphics[scale=0.4]{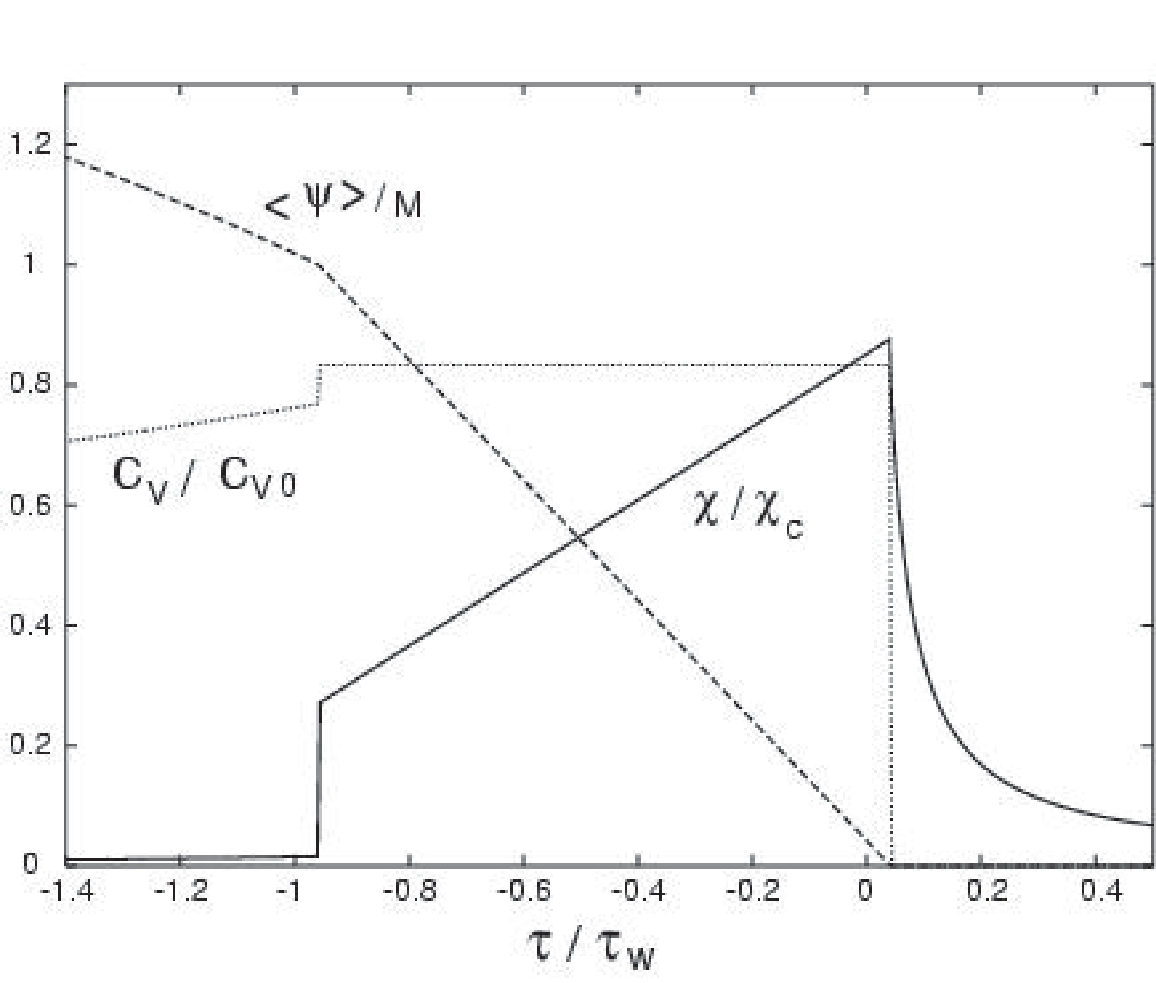}
 \caption{Normalized average order 
parameter $\av{\psi}/M$, 
normalized susceptibility $\chi/\chi_c$, 
and normalized specific heat $C_V/C_{V0}$  
versus $\tau/\tau_w$ for $\beta/\bar{u}=1.2$ 
in the limit $h\rightarrow 0$, 
where $C_{V0}= TA_0^2/2 {\bar u}$. 
}
\end{figure}

In the two phase states in the temperature window,  
the average order parameter is given by\cite{commen1} 
\be 
\av{\psi}=  \phi \psi_1+ (1-\phi)\psi_0,  
\en 
which is continuously 
connected to  the solution of Eq.(2.22) in 
the one phase sates outside the window region. 
See Fig. 4 for $\av{\psi}$ 
as a function of $\tau$ and $h$ 
at $\beta/{\bar u}=1.2$.  The effective  isothermal 
susceptibility $\chi= ({\p \av{\psi}}/{\p h})_\tau$ 
 is  calculated from  
\be 
\chi=  
(\psi_1-\psi_0)\frac{\p \phi}{\p h}+ 
\phi\frac{\p \psi_1}{\p h}+
(1-\phi)\frac{\p \psi_0}{\p h},  
\en 
where the derivatives are performed at fixed $\tau$. 
See Fig. 5 for $\chi$ as a function of $\tau$ and $h$ 
at $\beta/{\bar u}=1.2$. We can see that $\chi$ is discontinuous 
at the boundary of the window region. There is no critical 
divergence in  $\chi$ at the field-induced criticality 
attained.  In particular, as $h\rightarrow 0$, it behaves as 
\be 
\chi= (1- 3\phi/4+ 2vM^2/3\beta)/(vM^4/3),
\en 
where $vM^4/3$ is the value of $\tau_{cx}$ as $h\rightarrow 0$. 
For $\tau>\tau_{cx}$ we have $\chi=1/\tau$ at $h=0$. 
Figure 6 displays the behavior of $\chi$ on the axis 
 in the limit  $h\rightarrow 0$.

Next we consider the specific heat at constant volume 
$C_V= -T {\p^2 \av{f}}/{\p T^2} $ (per unit volume) 
arising from the spin degrees of freedom, 
where $h$ is fixed. In the two phase coexistence 
with $h<h_c$, we use Eqs.(3.3) and (3.16) to obtain 
\be 
C_V=  TA_0^2/{2\beta}, 
\en 
which is independent of $h$ 
even for  $h>0$.  In the one phase region, 
we have 
$C_V=TA_0^2\psi^2/(\tau+3{\bar u}\psi^2+5v\psi^4)$, 
where $\psi$ is determined by Eq.(2.19). 
In particular, at $h=0$, 
$C_V = 0$ for $\tau>\tau_{cx}$ 
and $C_V=TA_0^2/{2\sqrt{{\bar u}^2-4v\tau}}$ 
for  $\tau<\tau_{cx}-\tau_w$. 
In Fig. 6, we show $C_V$ versus $\tau$ at $h=0$.

\subsection{Correlation length and surface tension}

\begin{figure}[htbp]
 \includegraphics[scale=0.4]{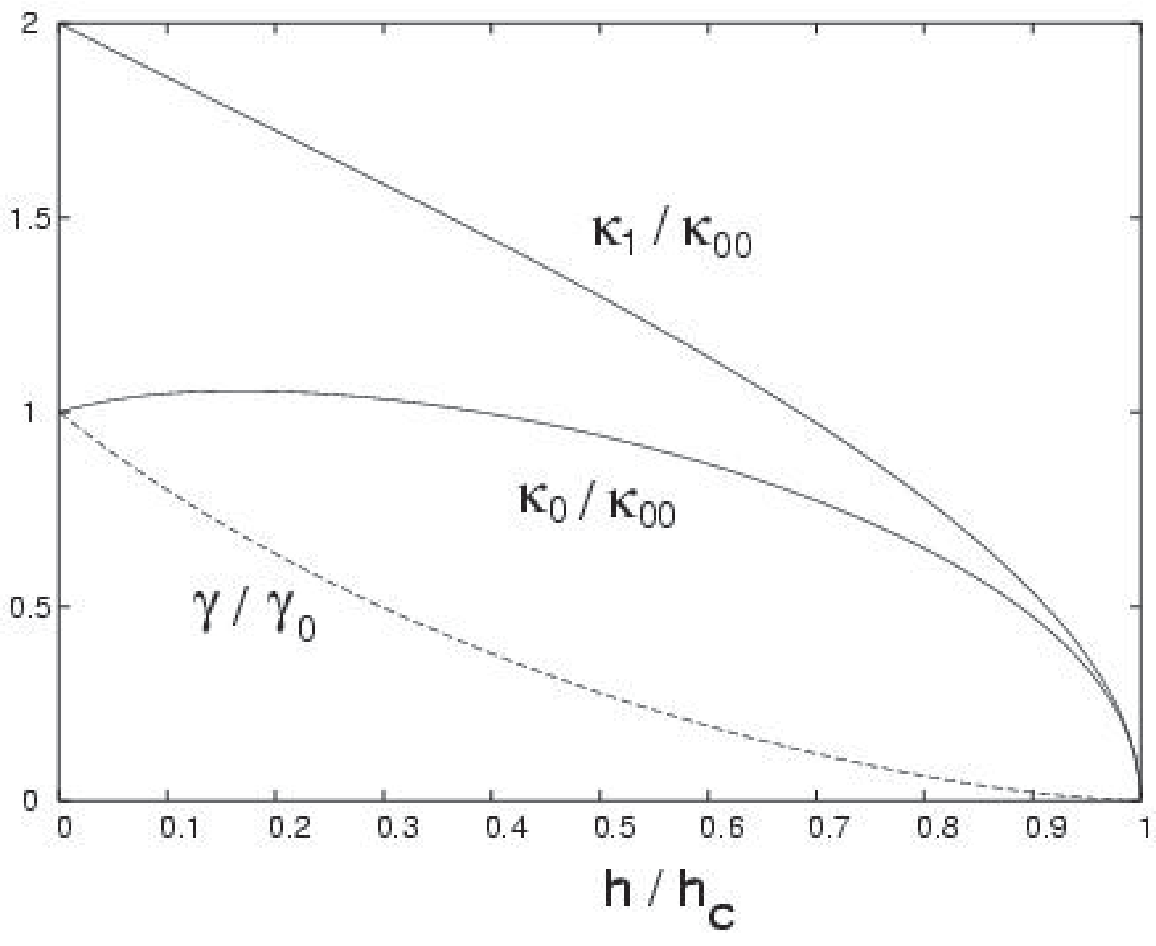}
 \caption{Inverse correlation lengths 
$\kappa_0$ and $\kappa_1$  versus $h/h_c$ 
in the coexisting two phases.  They are 
divided by $\kappa_{00}$ in Eq.(3.29). 
Normalized surface  tension $\gamma/\gamma_0$ 
is also shown, where $\gamma_0$ is in Eq.(3.36).
}
\end{figure}

Starting with  the first line of Eq.(2.27), we may calculate 
the inverse correlation lengths, 
$\kappa_0$ and  $\kappa_1$,  in the coexisting two phases 
with $\psi=\psi_0$ and $\psi_1$, respectively. 
With the aid of  Eqs.(3.6)-(3.8) some calculations yield  
\bea 
\kappa_0^2&=& \frac{v}{3C} (\psi_1-\psi_0)^2  (\psi_1+\psi_0)
 (\psi_1+4\psi_0), \\
\kappa_1^2&=& \frac{v}{3C} (\psi_1-\psi_0)^2  (\psi_1+\psi_0)
 (4\psi_1+\psi_0).
\ena 
As $h\rightarrow 0$, we have 
$\kappa_0 \rightarrow \kappa_{00}$ and  
$\kappa_1 \rightarrow 2\kappa_{00}$, where 
\be 
\kappa_{00}= (v/3C)^{1/2}M^2  
\en 
is the inverse correlation length in the disordered phase 
at $\tau=vM^4/3$ and $h=0$.  As $h\rightarrow h_c$, 
the inverse correlation lengths go to zero  as 
\be 
\kappa_0 \cong \kappa_1\cong (4/5)\kappa_{00}(1-h/h_c)^{1/2},
\en 
from Eq.(3.15). If the scattering amplitude is 
proportional to $S_k$ in Eq.(2.21), 
it grows near the critical point at long wavelengths. 
In Fig. 7, we plot 
$\kappa_0/\kappa_{00}$ and $\kappa_1/\kappa_{00}$ 
versus  $h/h_c$.    
It is worth noting that the inverse correlation length 
$\kappa$ in the one phase region also goes to zero 
at the criticality. In its vicinity, 
the relations  (2.19) and (2.22) in the one phase case give  
\be 
C\kappa^2\cong 
 (h-h_c)/\psi_c,
\en 
where the term linear in  $\tau-\tau_c$ vanishes.

We also calculate the surface tension $\gamma$ 
in the two phase coexistence.  We suppose  a one-dimensional 
interface profile $\psi=\psi(x)$ changing along the $x$ direction. 
 It changes from $\psi_0$ at $x=-\infty$ and 
to $\psi_1$ at $x=\infty$. From $\delta F/\delta\psi=0$, 
we obtain 
\be 
C\frac{d^2\psi}{dx^2}= f_0'(\psi) -\beta (\psi^2-\av{\psi^2})\psi.
\en 
We  integrate the above equation 
as $2\omega = {C}({d\psi}/{dx})^2$,  
where $\omega(\psi)$ is the grand potential, 
\be 
\omega=  f_0(\psi) -\frac{\beta }{4} 
(\psi^2-\av{\psi^2})^2 -C_0. 
\en 
From Eq.(3.6) the constant $C_0$ in the right hand side 
can be chosen such that $\omega$ vanishes at 
$x=\pm\infty$ or for both   $\psi=\psi_0$ and $\psi_1$. 
Some calculations yield \cite{Onukibook} 
\be 
\omega=   \frac{v}{3} 
(\psi-\psi_0)^2(\psi-\psi_1)^2
[(\psi+ \psi_0+\psi_1)^2 + \psi_0\psi_1], 
\en
which turns out to be independent of $\tau$. 
The surface tension $\gamma$ is a function of $h$ only. 
It is   of the form, 
\bea 
\gamma&=& \int_{-\infty}^\infty
dx [ \omega+ C(d\psi/dx)^2/2]\nonumber\\  
 &=& \int_{\psi_0}^{\psi_1}d\psi 
\sqrt{2C\omega(\psi)} .
\ena 
In the limit  $h\rightarrow 0$  it becomes  
\be 
\gamma_0= \lim_{h\rightarrow 0}\gamma= (vC/24)^{1/2}M^4.
\en 
On the other hand, as $h\rightarrow h_c$, $\omega$ in (3.35) 
behaves as  
$\omega \cong |u|(\psi-\psi_0)^2(\psi-\psi_1)^2/2$ so that  
\be 
\gamma/\gamma_0 \cong  (32/375)(1-h/h_c)^{3/2},
\en 
which rapidly decreases near the criticality. 
See Fig. 7, where $\gamma/\gamma_0$ is plotted.

\section{Numerical Results}
\setcounter{equation}{0}
We numerically study the dynamics of our model. 
We may  demonstrate 
the validity of our equilibrium theory in steady states 
attained at long times. 
In our system $\psi$ is a nonconserved 
variable obeying  the  relaxation equation, 
\be 
\frac{\p}{\p t}\psi=-L_0\frac{\delta F}{\delta\psi}, 
\en 
where ${\delta F}/{\delta\psi}$ is given in Eq.(2.14) 
and $L_0$ is a constant. 
We  integrated the above equation in 2D 
under the periodic boundary condition.  
We assume $\bar{u}>0$ and $\beta/{\bar u}=1.5$. 
Then,  for  $h=0$,  our theory predicts 
$\psi_1 =0.612M_0$,  $\psi_0=0$, $\tau_{cx}/\tau_0= 
0.047$, $(\tau_{cx}-\tau_w)/\tau_0=-0.516$, 
$\kappa_0 \ell =0.354$, and $\kappa_1\ell =0.596$.@ 
These values will be compared with 
those from our simulations.

\subsection{Isotropic elasticity}

\begin{figure}[t]
 \includegraphics[scale=0.4]{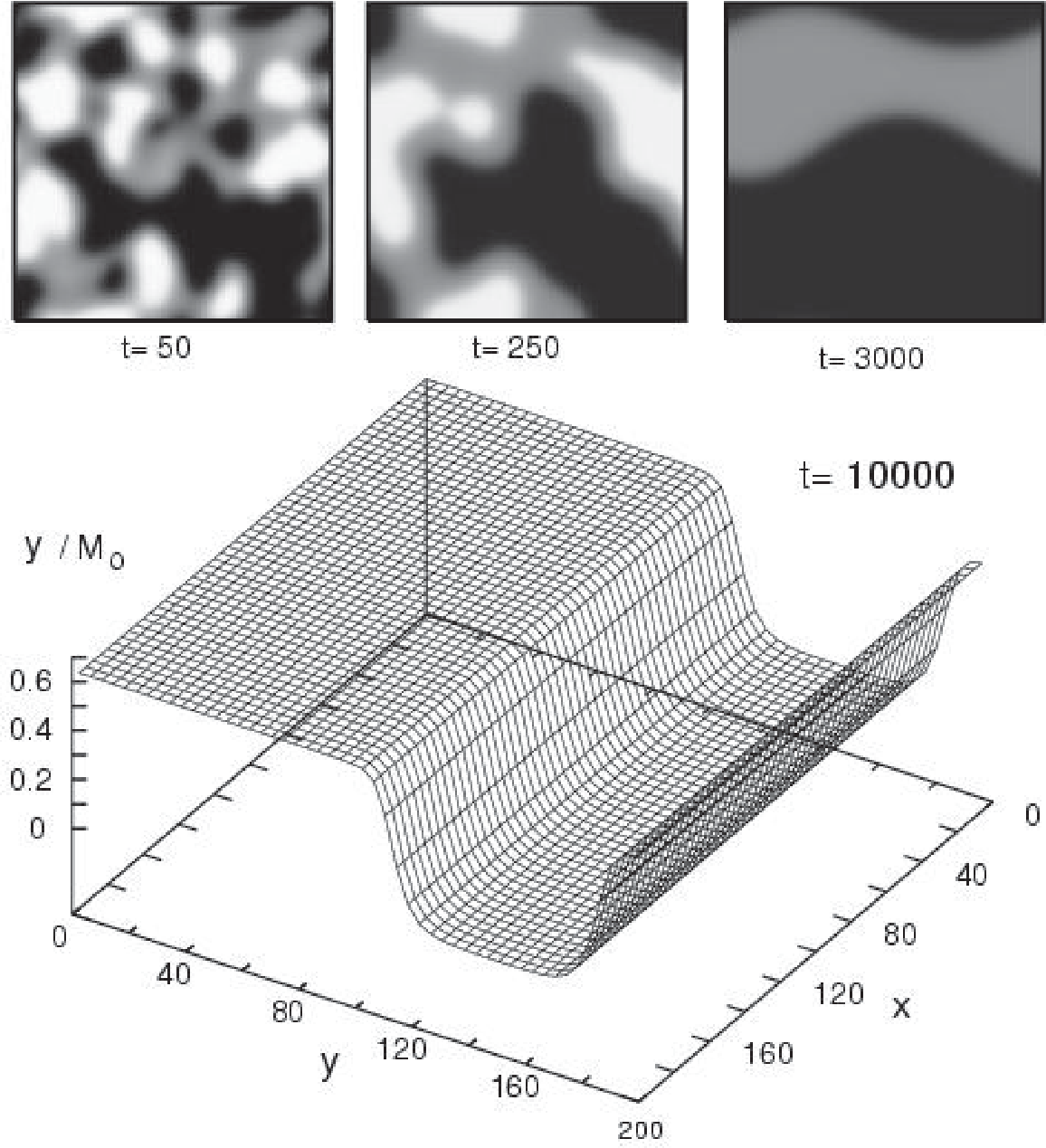}
 \caption{Time evolution of $\psi$ 
after changing $\tau$ from 0 to $-0.3\tau_0$ at $t=0$ 
(upper panel) and steady  profile of $\psi/M_0$ obtained at 
$t=10^4$ (lower panel) 
for $\beta/{\bar u}=1.5$ and $h=0$  in  isotropic elasticity. 
Here $M_0$ is in Eq.(4.2) 
and space and time are 
 measured in units of $\ell$ and $t_0$  in Eq.(4.3). 
In the initial stage three regions with 
$\psi \cong \psi_1$ (black), 
$\psi \cong -\psi_1$ (white), 
and $\psi \cong 0$ (gray) emerged, but 
in the final stage $t\gs 10^3$ 
the variant with $\psi \cong -\psi_1$ 
 disappeared  here. 
}
\end{figure}

We first assume  the isotropic elasticity. 
We measure $\tau$,   $h$, and $\psi$ in units of 
$\tau_0$, $h_0$, and $M_0$, respectively, where  
\be 
\tau_0= {\bar u}^2/v, \quad h_0= v({\bar u}/v)^{5/2}, \quad 
M_0= ({\bar u}/v)^{1/2}.
\en 
Here $M/M_0=[{3(\beta/{\bar u}-1)}/4]^{1/2}$  from Eq.(3.9). 
Units of space and time are 
\be 
t_0= L_0\tau_0, \quad \ell= (C/\tau_0)^{1/2}.
\en 
The scaled time $t_0^{-1}t$ and 
the scaled space position $\ell^{-1}{\bi r}$ are  
simply written as $t$ and $\bi r$ to avoid cumbersome notation. 
The system size is $200\times 200$ and  the mesh length 
is  $\ell$, so the system length is $200\ell$. 
In terms of the scaled order parameter 
$\Psi=\psi/M_0$,  Eq.(4.1) is rewritten as 
\be 
\frac{\p\Psi}{\p t}= \bigg[\nabla^2 -\frac{\tau}{\tau_0}-
\Psi^2-\Psi^4 + \frac{\beta}{\bar u}
(\Psi^2-\av{\Psi^2}) \bigg]\Psi + \frac{h}{h_0} . 
\en
As the initial condition at $t=0$, 
$\Psi$ at each lattice point 
consists of a homogeneous constant  
and  a random number in the range  
$[-0.01,0.01]$.

\begin{figure}[bt]
 \includegraphics[scale=0.4]{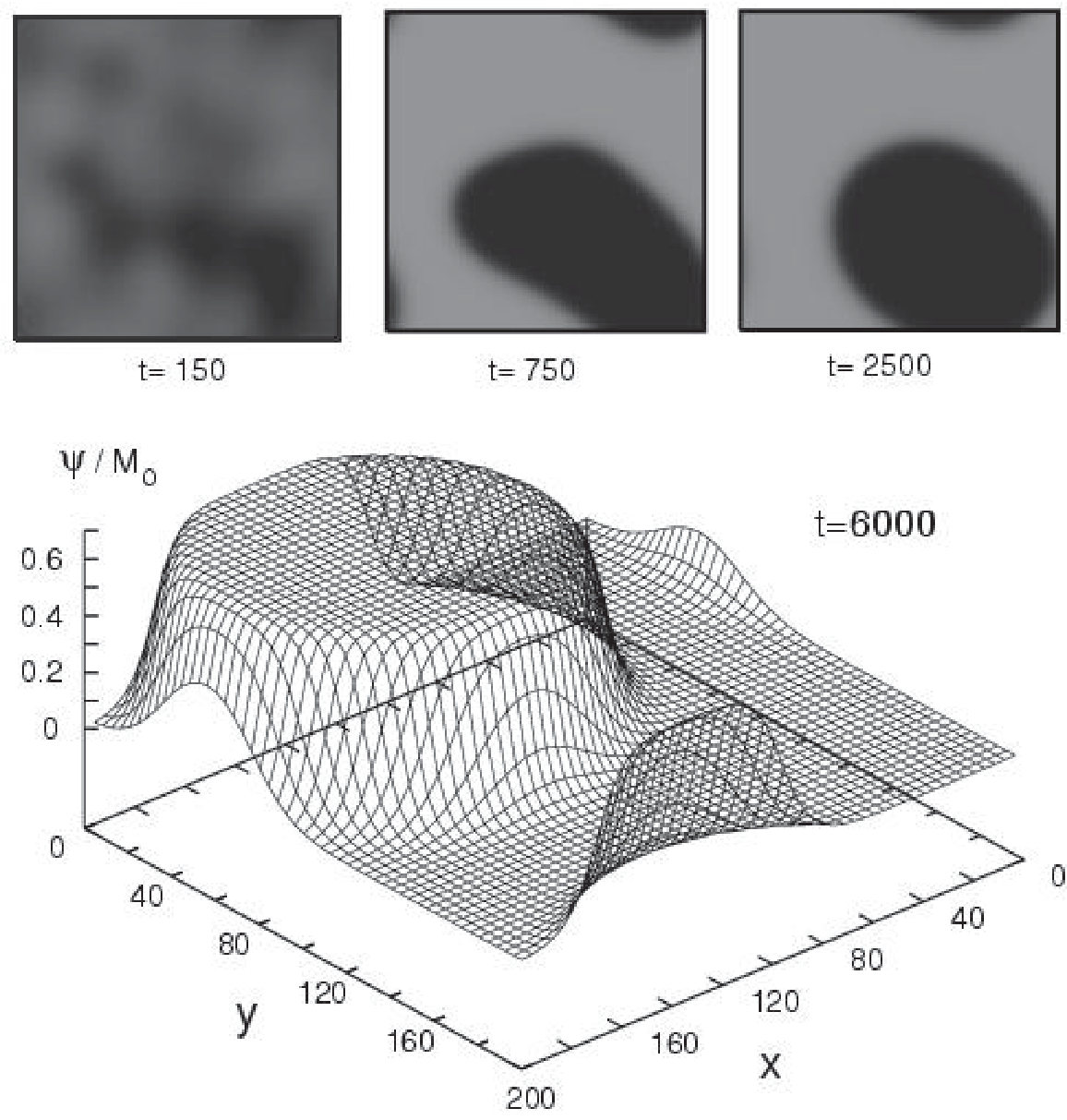}
 \caption{Time evolution of $\psi$ 
after changing $\tau$ from $-\tau_0$ to $-0.15\tau_0$ at $t=0$ 
(upper panel) and steady profile of $\psi/M_0$ obtained at 
$t=6\times 10^3$ (lower panel) 
for $\beta/{\bar u}=1.5$ and $h=0$  in  isotropic elasticity. 
In the phase ordering,  ordered regions with 
$\psi \cong \psi_1$ (black) 
and  disordered regions   (gray) emerged. 
A circular ordered domain remained at long times in this run.}
\end{figure}

In Fig. 8, we show  
the phase ordering process from 
a  disordered state  to a coexisting 
state.   At $t=0$,  $\Psi$ 
 was a random number. 
For $t>0$ we lowered  $\tau$ from 0 
to $-0.3\tau_0$  to induce phase ordering. From  our theory,  
 this final $\tau$ is  in the 
coexisting window 
 $[\tau_{cx}-\tau_w,\tau_{cx}]$ 
and the predicted 
average order parameter is 
$0.378M_0$ with  $\phi= 0.617$. 
Since $h=0$ and $\av{\psi}=0$ at $t=0$, 
the two variants with $\psi= \pm \psi_1$ 
appeared  in the early stage, but the ordered domains 
with $\psi\cong  - \psi_1$ disappeared in this run 
when  the domain size became of the order of  the system size. 
(In  other runs the variant with $\psi\cong 
\psi_1$ disappeared as well.) 
In the steady two phase coexistence 
at $t=10^4$ (lower panel in Fig. 8) interfaces are horizontal (parallel 
to the $x$ axis), where  
 $\psi = 0.612M_0$ in the ordered phase  
and $\av{\psi}=0.397M_0$. 
The former  coincides  
with the predicted value,   while 
the latter is slightly larger than predicted.

In Fig. 9,   we show 
the phase ordering process  from 
a  one phase state at $\tau=-\tau_0$  to a coexisting 
state at $\tau=-0.15\tau_0$  at $h=0$.  That is, 
at $t=0$,  $\Psi$ 
 was the sum of  the equilibrium one phase 
value $0.786$ determined by Eq.(2.24) and 
 a random number. 
The final $\tau$ here is higher 
 than the lower instability value 
$-0.313\tau_0$  in   Eq.(2.23). 
Hence  phase ordering should take place  
into a coexisting state where  
$\phi=0.350M_0$ and $\av{\psi}= 0.214 M_0$ are predicted. 
In  the simulation,   regions of 
the disordered phase appeared, while $\psi$ in the 
 ordered phase changed to  $\psi\cong  \psi_1$. 
In the steady two phase coexistence 
at $t=6\times 10^3$ (lower panel in Fig. 9), 
a circular ordered domain was  realized. 
There,  we find 
 $\psi = 0.594M_0$ in the domain   
 and $\av{\psi}=0.232M_0$. These 
values are only slightly different from  those  
  predicted.

\begin{figure}[bt]
 \includegraphics[scale=0.4]{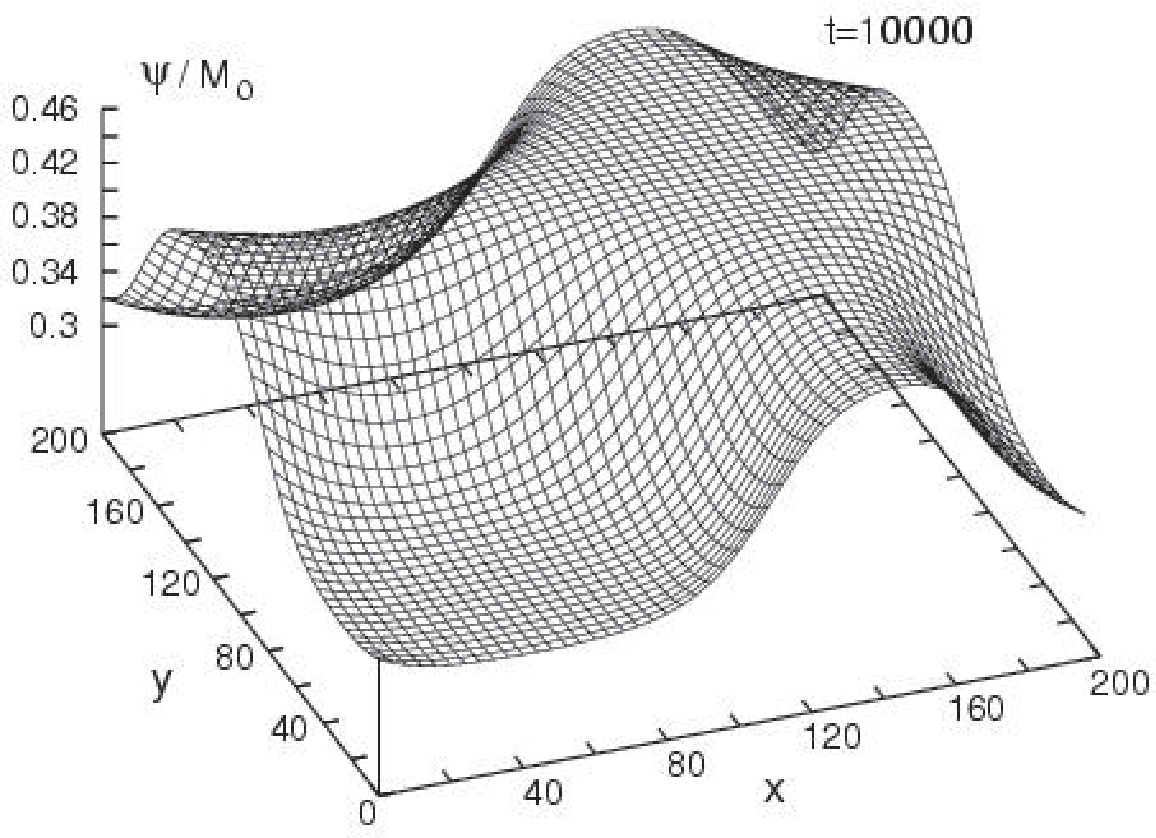}
 \caption{Steady profile of $\psi/M_0$ 
in two phase coexistence obtained 
at $t= 10^4$   in  isotropic elasticity,  where   
$\beta/{\bar u}=1.5$, $\tau= -0.13\tau_0$,  and $h=0.9h_c$. 
Here the system is close to the field-induced critical point 
and the interface region is broadened.}
\end{figure}

In Fig. 10, we present a steady profile of 
$\psi$ at  $h=0.9 h_c$ and $\tau= -0.13\tau_0$, 
where the system is  close to the critical point in Eqs.(3.12)-(3.14) 
and the interface thickness is much widened. 
For $\beta/{\bar u}=1.5$ and at this field, 
  our theory gives 
$\psi_1/M_0=0.461$,  $\psi_0/M_0=0.304$, 
$\tau_{cx}/\tau_0= 
-0.032$, $(\tau_{cx}-\tau_w)/\tau_0=-0.212$, 
$\kappa_0 \ell =0.149$, and $\kappa_1\ell =0.164$.
For the $\tau$ adopted,  we predict 
$\phi= 0.543$  and $\av{\psi}= 0.389M_0$.   
In the simulation,  
the maximum  and the minimum 
of $\psi$ are  $0.457M_0$ 
and  $0.302M_0$, respectively. These values  are very close to 
the theoretical values of $\psi_1$ and $\psi_0$. 
Furthermore, the observed average 
$\av{\psi}= 0.392M_0$  is also close to its  
theoretical average, though the interface regions are  very wide here.

\subsection{Cubic  elasticity}

Next we integrate Eq.(4.1) in 2D on a cell of $256\times 256$ 
assuming the cubic elasticity 
with   $C_{11}-C_{12}=C_{44}=K$,  where $K=(C_{11}+C_{12})/2$. 
Then $\xi_a=-1$ from Eq.(2.11) 
and the softest directions are $[10]$ and $[01]$. 
As  in the isotropic case,  
space and time are measured in units of $\ell$ and 
$t_0$ in Eq.(4.3) and we set $\beta=2\alpha^2/C_{11}= 1.5 {\bar u}>0$. 
The mesh size of integration is  $\ell$. 
In terms of the scaled order 
parameter $\Psi=\psi/\psi_0$, 
the dynamic equation in the 2D cubic 
case is written as \cite{Onukibook,comment,Onuki_cubic,Nie}
\be 
\frac{\p\Psi}{\p t}= \bigg[\nabla^2 -\frac{\tau}{\tau_0}-
\Psi^2-\Psi^4 + \frac{\beta}{\bar u}
G  \bigg]\Psi + \frac{h}{h_0} . 
\en
From Eqs.(2.8) and (2.14) 
we express $G({\bi r})$  in the Fourier expansion, 
\be 
 G({\bi r})= \frac{1}{w(0)}    
\sum_{{\bi k}} {w(\theta)}
\Phi_{\bi k}  e^{i{\bi k}\cdot{\bi r}},  
\en
 where   $\Phi_{\bi k}$ is the Fourier component of 
$\Phi= \Psi^2-\av{\Psi^2}$  and 
 $w(\hat{\bi k})=w(\theta)$  in Eq.(2.13) 
depends on the angle $\theta$ defined by  
$\cos\theta=k_x/k$.

\begin{figure}[bt]
 \includegraphics[scale=0.4]{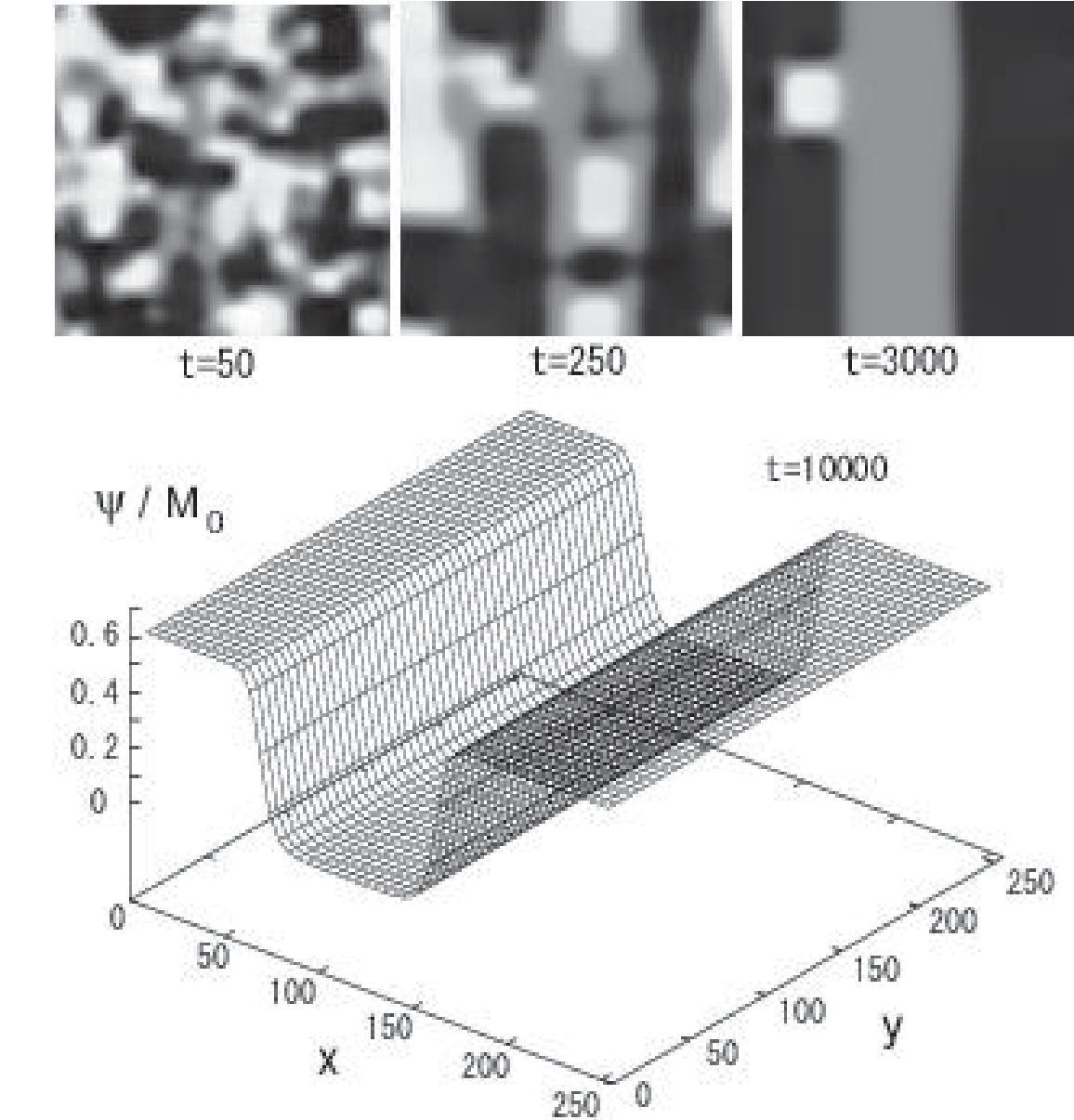}
 \caption{
Time evolution of $\psi$ 
after changing $\tau$ from 0 to $-0.3\tau_0$ at $t=0$ 
(upper panel) and final steady profile of $\psi/M_0$ obtained at 
$t=10^4$ (lower panel)   in  cubic  elasticity 
for $\beta/{\bar u}=1.5$ and $h=0$. 
As in Fig.8,  three regions with 
$\psi \cong \psi_1$ (black), 
$\psi \cong -\psi_1$ (white), 
and $\psi \cong 0$ (gray) emerged 
in the initial stage. Interfaces tend to be parallel to the $x$ or $y$ 
axis. }
\end{figure}
\begin{figure}[bt]
 \includegraphics[scale=0.4]{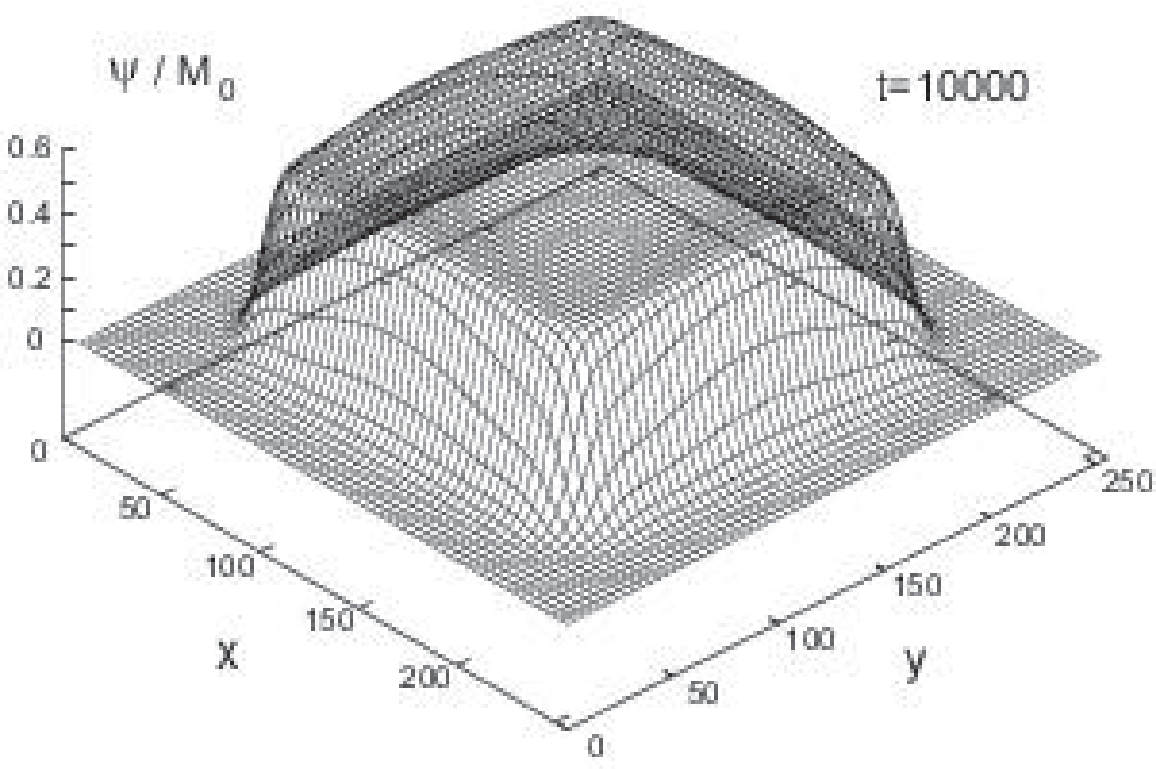}
 \caption{ Steady square profile of 
$\psi/M_0$ in two phase coexistence obtained 
at $t= 10^4$  in cubic elasticity,  where   
$\beta/{\bar u}=1.5$, $\tau= -0.15\tau_0$,  
and $h=0$.} 
\end{figure}
\begin{figure}[bt]
\includegraphics[scale=0.4]{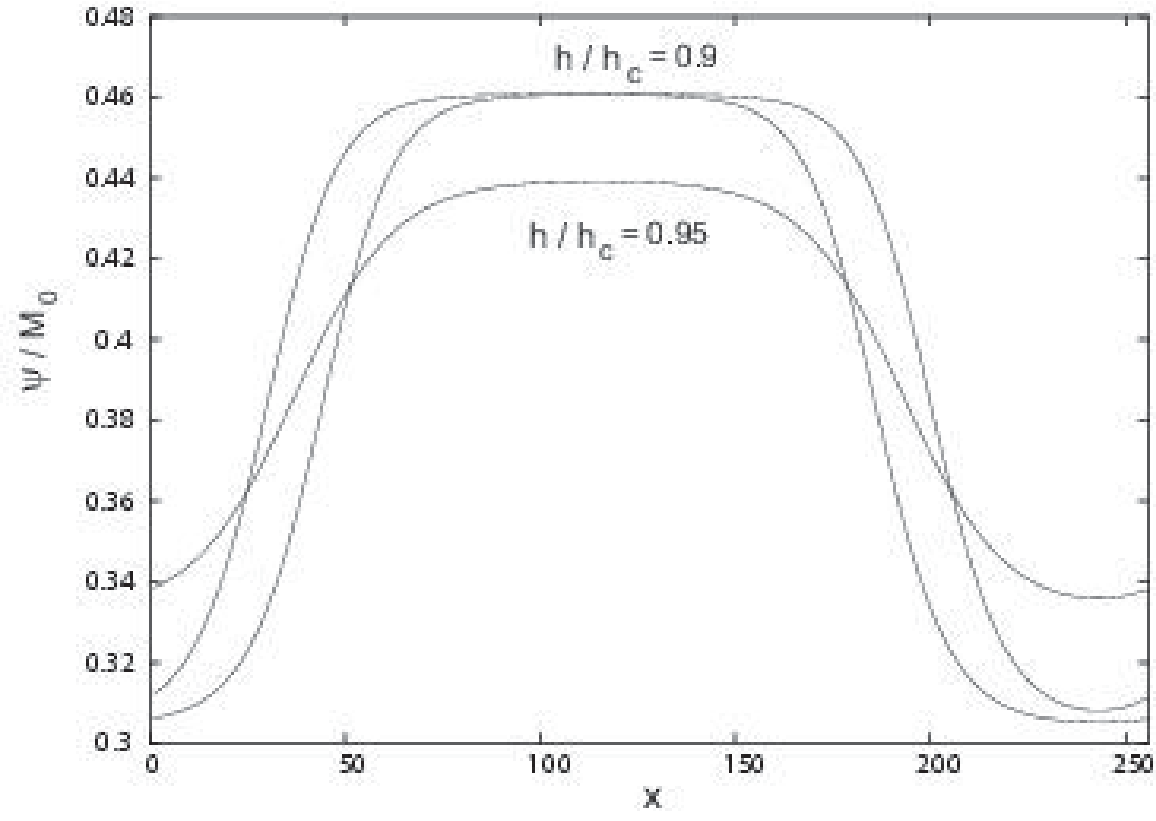}
 \caption{Steady,  one-dimensional 
curves  of $\psi/M_0$ 
in two phase coexistence with $\beta/{\bar u}=1.5$ 
near the field-induced critical point in cubic elasticity. 
Here  $\tau= -0.13\tau_0$  and $-0.15\tau_0$ 
for the two curves of $h=0.9h_c$, 
while $\tau=-0.13\tau_0$  for  $h=0.95h_c$. 
The more ordered region expands with lowering $\tau$ at fixed $h$.}
\end{figure}

In Fig. 11, 
we lowered $\tau$ from 0 to $-0.3\tau_0$ at $h=0$ as in Fig. 8. 
Here the anisotropy of the domain structure arises from the 
angle dependence of $w(\theta)$ in Eq.(4.6). 
In the steady state in the lower panel, the maximum of $\psi$
is $0.613M_0$ and the average $\av{\psi}$ 
is $0.393M_0$,  in close agreement with the predicted 
values and those in Fig. 8.

In Fig. 12, we show a steady  profile of $\psi$ 
for $\tau=-0.13\tau_0$ and $h=0$ as in Fig. 9. 
Here a square ordered domain is embedded 
in a disordered region in equilibrium. 
In the figure, the maximum and the average 
of $\psi$ are  $0.586M_0$ 
and  $0.260M_0$, respectively. 
The former is slightly smaller than 
the predicted value $0.612M_0$, while the latter 
is considerably larger than the predicted value 
 $0.214M_0$.

In Fig. 13, we show one-dimensional steady profiles 
changing along the $x$ axis near the critical point. 
The maximum,  the minimum, 
and the average   of $\psi$ are 
$(0.461, 0.305, 0.392)$ for $h/h_c=0.9$ and $\tau/\tau_0= 
-0.13$, 
$(0.461, 0.308, 0.409)$ for $h/h_c=0.9$ and $\tau/\tau_0= 
-0.15$,  and 
$(0.439, 0.336, 0.398)$ 
for $h/h_c=0.95$ and $\tau/\tau_0= 
-0.13$.  These values closely agree with those  from our theory.
In these one-dimensional cases, 
the profiles   coincide with   those 
in the isotropic case.

\section{Summary  and concluding remarks} 

We have examined 
the phase transition behavior of compressible Ising 
models at fixed volume in the mean field theory.  
In our model  the  order parameter $\psi$ 
is isotropically coupled to the dilation strain $e_1$ 
as $\psi^2e_1$ in the free energy, 
which is the simplest case. 
Nevertheless,   complicated 
phase behavior follows  at constant volume.  
We summarize our main results.\\ 
(i) We have found two phase coexistence 
in a closed region in the $\tau$-$h$ plane  as in Fig. 3. 
The  coexistence region appears under the condition ${\bar u}<\beta$ 
given in Eq.(3.1).  If   $\bar{u}>0$  
and   $\beta$ is not large, it can be  satisfied 
 near the tricritical point.    If  ${\bar u}<0$, 
it can occur even away from the tricritical point. \\ 
(ii) The order parameter  values 
in the two phases, $\psi_1$ and $\psi_0$, 
  are determined by $h$ only and is independent of 
$\tau$   as in Fig. 2. 
The average order parameter 
$\av{\psi}=\phi \psi_1+(1-\phi)\psi_0$ 
is increased  smoothly as $\tau$ is decreased  in the 
 window region $\tau_{cx}-\tau_w<\tau<\tau_{cx}$ 
for $h<h_c$, since the volume fraction $\phi$ 
depends on $\tau$ as in Eq.(3.16). 
The average order parameter $\av{\psi}$ 
and the susceptibility $\chi=\p\av{\psi}/\p h$ 
are displayed in Figs. 4-6.  
The specific heat $C_V$ is a constant  in two phase 
coexistence  as in Eq.(3.26).\\
(iii)   
At the field-induced critical point 
$h=h_c$ and $\tau=\tau_c$, 
the correlation length $1/\kappa$ grow and the 
 surface tension $\gamma$ 
goes to zero as in Fig. 7,  while $\chi$ does not diverge.\\
(iv)  
We have integrated the dynamic equation, which is 
 Eq.(4.4) for the isotropic elasticity 
and  Eq.(4.5) for the cubic elasticity. 
A  change of $\tau$ from the one phase region 
into the unstable 
region induces   phase ordering as 
illustrated in  Figs. 8-13. 
It can occur  with 
decreasing $\tau$ as in Figs. 8 and 11 
and with increasing $\tau$ as in Fig. 9.  
In the final two phase states, 
the  values of $\psi$ and its space  average  
  closely  agree  with 
the theoretical values.

We make some further remarks.\\
(i) 
At constant pressure, two phase coexistence 
occurs only on a line in the $\tau$-$h$ plane 
as in the rigid lattice case, but  phase separation can be 
much affected by the elastic 
coupling (see the appendix) \cite{Little}. It is worth noting that the 
transition depends on the sample shape 
in hydrogen-metal systems at constant pressure 
\cite{Onukibook,Wagner}, where the proton concentration 
is linearly coupled to the dilation \cite{comment}.\\ 
(ii) We mention   Monte Carlo simulations on  a binary alloy   
by Landau's  group \cite{La1,La3,La4}. 
 They assumed   that  a mixture undergoing 
unmixing corresponds to ferromagnets and 
that forming a superstructure  to 
antiferromagnets. In these  cases, 
 different results followed in the fixed volume 
and fixed pressure conditions.  However,  the unmixing transition 
in the presence of the  size difference\cite{Onukibook} is 
 not isomorphic  to the ferromagnetic transition. 
In the former   the linear coupling 
\cite{comment} appears between the concentration 
$c$ and $e_1$ in the form $\psi e_1$, 
while in the latter the exchange interaction 
does not break  the invariance of  $\psi\rightarrow -\psi$ 
and the elastic  coupling is quadratic as $\psi^2 e_1$. 
At present we cannot compare 
  our theory  and  their  simulations. 
\\ 
(iii) Yamada and Takakura numerically solved 
a time-dependent Ginzburg-Landau model 
for an  order parameter and a strain 
in  one dimension.  They found appearance  of a disordered 
region in a lamellar ordered region \cite{Yamada}. 
 Their finding is consistent with  our theory.\\  
(iii) In real metamagnets,  
there is  no  field  conjugate to the antiferromagnetic order 
 and the tricriticality has been realized 
by changing  magnetic field or  hydrostatic pressure.   
At fixed volume, our theory predicts 
 two phase coexistence 
 in a temperature window near the tricritical point 
and near the line of first order phase transition. 
From Eq.(3.21) the width of the  window 
sensitively depends on the coupling constant $\alpha$ as  
$\tau_w/A_0=3\beta (\beta-{\bar u})^2/4vA_0$, where $A_0$ is 
the coefficient in   Eq.(2.3)  and $\beta=2\alpha^2/K$.\\ 
(iv)  
In our mean field theory,  
we have neglected the renormalization  effect near 
the critical point, which can be intriguing 
in the presence of the cubic elastic anisotropy 
 \cite{Lubensky,Halperin}. 
It should be  further studied together with 
the influence  of the global elastic constraint
studied in this work.\\
(v) 
We should generalize our theory to  more complex systems. 
At the ferroelectric transition \cite{Gammon}, 
 the polarization vector is coupled to the 
strains. In binary alloys, phase separation and an order-disorder 
phase transition can take place simultaneously 
\cite{Onukibook}, where 
the concentration $c$ and the structural order 
parameter $\psi$ are both  coupled to 
$e_1$  in the form $(\alpha_1c+\alpha_2\psi^2)e_1$  
in the free energy \cite{Sagui94}. 
There can also be a number of  
anisotropic elastic couplings between the order parameter 
and  the tetragonal or   
 shear strain.  We will soon report on phase transition  
including a Jahn-Teller coupling 
 \cite{JT}.\\

\begin{acknowledgments}  
We would like to thank 
B. D$\ddot{\rm u}$nweg for informative correspondence. 
This work was supported by Grants in Aid for Scientific Research and for
the 21st Century COE project (Center for Diversity and Universality in
Physics) from the Ministry of Education, Culture, Sports, Science and 
Technology of Japan.   

\end{acknowledgments}

\vspace{2mm} 
{\bf Appendix  A: Fixed pressure  condition}\\
\setcounter{equation}{0}
\renewcommand{\theequation}{A.\arabic{equation}}

We here eliminate the  elastic field at fixed pressure 
\cite{Larkin,Sak,La3}. 
Under  isotropic applied stress, 
we assume an isotropic average dilation change 
$\av{e_1}$ caused by the order parameter change. 
The average stress should be unchanged from 
that in the reference state, so we require 
 $\av{\sigma_{ij}}=0$ in Eq.(2.6) to obtain 
\be 
\av{e_1}= -\alpha\av{\psi^2}/K, 
\en 
in terms of the bulk modulus $K$. 
We  impose the periodic boundary condition 
on the  deviation, 
$
\delta u_i= u_i - \av{e_1}x_i/d,  
$  
whose Fourier component 
can be expressed in terms of $\varphi_{\bi k}$ 
in the same form as that of $u_i$ 
in the fixed volume case. The 
 free energy consists of $F$ in 
Eq.(2.15) and 
\be 
\Delta F= -V{\alpha^2} \av{\psi^2}^2/2K.
\en 
The total free energy $F'=F+\Delta F$ is written as  
\be
{F}' =  \int d{\bi r}\bigg [ f 
+\frac{C}{2} |\nabla\psi|^2 + 
\frac{B}{4} (\psi^2 -\av{\psi^2})^2  \bigg] ,
\en  
where $f=f_0- {\alpha}^2\psi^4/2K$ 
and $B$ is a positive coefficient,  
\be 
B = 2\alpha^2/K - 2w_M.
\en 
Here  $w_M$ is given by Eq.(2.17). 
The positivity of $B$ arises from 
$C_{11}-C_{12}>0$ and $C_{44}>0$. 
The one phase ordered states are determined by $f$. 
The  same  form of the free energy was derived by Littlewood and 
Chandra \cite{Little} for  BaTiO$_3$,  who argued that 
the term proportional to $B$ 
can much decrease the nucleation rate 
from the paraelectric to ferroelectric state.  
In our problem, we draw  the following conclusion in the mean field 
theory. In the fixed pressure condition,  there can  be 
two phase coexistence only on  a first-order coexistence 
line   in the $\tau$-$h$ plane. In fact,  $\av{f}$ in Eq.(3.3) 
would be minimized for $\phi=0$ or 1 outside 
the coexistence curve 
if positive $\beta$ were  replaced by negative $-B$.


\begin{references}
\bibitem{comment} In binary mixtures \cite{Onuki_cubic,Nie}, 
the concentration $c$ is coupled 
in the form $\alpha  c e_1$ in the free energy density 
when the atomic sizes of the two components 
are different \cite{Onukibook}.  There,  
  $\varphi_{\bi k}$ in  Eqs.(2.8) and (2.12) 
is replaced by the Fourier 
component $c_{\bi k}$ of the concentration. 
The concentration dependence of the elastic moduli 
is relevent here, 
leading to a first order phase transition into 
a glassy ordered  state \cite{Onukibook}.
\bibitem{comment2} For the free energy density (3.22) 
the first order phase transition line is expressed 
as $\tau= v(4M^4- 6M^2q^2+5q^4)/9$ in terms of $q=\psi_1+\psi_0$, 
where $q$ is related to $h$ by Eq.(3.11)  and  
$\psi_1-\psi_0= [{(8M^2-5q^2)/3}]^{1/2}$.
\bibitem{commen1} At  $h=0$, 
 the two variants with $\psi=\pm \psi_1$ 
may be present  in  large systems, while 
only one of them    survived in 
our simulations in Figs. 8 and 11.   
Thus we assume $h>0$ in Eq.(3.23) and take the limit 
$h\rightarrow 0$ in Fig. 6.
\end{references}
\end{document}